\documentclass[conference,a4paper]{APSIPA2020}
\usepackage{multirow}
\usepackage[dvips]{graphicx}
\usepackage{amsmath}
\usepackage[psamsfonts]{amssymb}
\usepackage{amsxtra}
\usepackage{threeparttable}

\usepackage{cite,balance}
\usepackage{epsfig,amssymb,amsmath,multirow}

\begin{document}

\title{HLT-NUS Submission for 2019 NIST Multimedia Speaker Recognition Evaluation}

\author{%
\authorblockN{%
Rohan Kumar Das, Ruijie Tao, Jichen Yang, Wei Rao, Cheng Yu and Haizhou Li
}
%
%
Department of Electrical and Computer Engineering,\\
National University of Singapore, Singapore\\
E-mail: \{rohankd, eleyji, haizhou.li\}@nus.edu.sg, ruijie.tao@u.nus.edu
}

\maketitle
\thispagestyle{empty}

\begin{abstract}

This work describes the speaker verification system developed by Human Language Technology Laboratory, National University of Singapore (HLT-NUS) for 2019 NIST Multimedia Speaker Recognition Evaluation (SRE). The multimedia research has gained attention to a wide range of applications and speaker recognition is no exception to it. In contrast to the previous NIST SREs, the latest edition focuses on a multimedia track to recognize speakers with both audio and visual information. We developed separate systems for audio and visual inputs followed by a score level fusion of the systems from the two modalities to collectively use their information. The audio systems are based on x-vector based speaker embedding, whereas the face recognition systems are based on ResNet and InsightFace based face embeddings. With post evaluation studies and refinements, we obtain an equal error rate (EER) of 0.88\% and an actual detection cost function (actDCF) of 0.026 on the evaluation set of 2019 NIST multimedia SRE corpus.

\end{abstract}

\section{Introduction}
\label{sec:intro}

Speaker recognition is one of the widely studied areas in speech processing that has witnessed success over the last two decades~\cite{Tomi,hansen_review}. Such success has made possible to deploy the technologies from laboratory to real-world applications in the recent years~\cite{sv_debut,smart_home,ijst_deb,ncciv2014,Das2017,APSIPA_realism,SpeechMarker}. The National Institutes of Standards and Technology (NIST) runs a speaker recognition evaluation (SRE\footnotemark[1]) to benchmark the robustness of various systems in adverse and realistic conditions~\cite{greenberg2020two}. Some of those adverse conditions included in the past few editions are short utterance, noisy environment, channel mismatch, language mismatch and multi-speaker test scenario~\cite{greenberg2020two}. The top performing teams in every editions show the scope towards addressing these issues by development of robust systems~\cite{Villalba2019,I4U2019,Torres-Carrasquillo2017,rkd_indigo2017}.

The latest edition 2019 NIST SRE runs on two tracks. The first track focuses on addressing the domain mismatch in terms of language, similar to that of the past two editions in 2016 and 2018~\cite{SadjadiCTS2019}. The only difference being removal of fixed set training condition and to observe the specific gain achieved by open set training condition. On the other hand, the second track sets out an entirely new direction to explore the multimedia speaker recognition for the first time as a part of SRE~\cite{sadjadiSRE2019}. It aims to recognize a speaker's identify collectively with both audio and visual cues. This paper focuses on our participation to 2019 NIST multimedia SRE and the developed systems for the challenge with their associated results.   

\footnotetext[1]{https://www.nist.gov/itl/iad/mig/speaker-recognition}

The research on processing multimedia for various applications have witnessed breakthrough in the recent years~\cite{baltruvsaitis2018multimodal}. It has been successfully explored for applications such as speech separation~\cite{ephrat2018looking}, speaker diarization~\cite{hoover2017putting} and automatic speaker recognition~\cite{noda2015audio}. While there are few works~\cite{AV1999,AV2003,AV2007,AV2016} exploring audio-visual multimedia information in the context of speaker recognition, they are not comparable to one another due to lack of benchmark data in the past. Further, the very recent explorations for benchmarking latest developments focus dominantly either on speaker recognition or face recognition, which is independent of multimedia processing. In this regard, the latest 2019 NIST multimedia SRE plays a pivotal role to have a common platform for evaluating performance of audio-visual speaker recognition systems.     

Both speaker recognition and face recognition systems have evolved over the time to reach the current state-of-the-art systems. In this work, we use x-vector based speaker embedding for speaker recognition studies~\cite{xvectors}. A previous study~\cite{NEC-TT2018CSL} showed that mixed bandwidth systems help to achieve improved results for speaker recognition. Therefore, we resample the audio to build narrowband and wideband audio data based separate systems. Additionally, we focus on speech enhancement aspect as well to process the data from real-world scenario in the challenge. On the other hand, the visual system is based on ResNet-101 model for performing face recognition~\cite{megapixels}. Post evaluation, we also developed more advanced InsightFace based system for face recognition studies~\cite{deng2019arcface}. Finally, a score level fusion is performed for the developed audio and visual systems to use audio-visual information for the 2019 NIST multimedia SRE collectively.

The rest of the paper is organized as follows. Section~\ref{secii} describes the details of the audio systems developed for speaker recognition studies. In Section~\ref{seciii}, the visual systems developed for face recognition studies are described. The results and analysis for the submitted and post evaluation systems are reported in Section~\ref{seciv}. Finally, the paper is concluded in Section~\ref{conc}.

\section{Audio Systems}
\label{secii}

This section describes the details of audio systems developed as a part of 2019 NIST multimedia SRE. The audio systems follow the x-vector based speaker modeling~\cite{xvectors}. We focus on two different aspects while developing the audio systems. The first aspect deals with development of mixed bandwidth (narrowband and wideband) systems, whereas the second one focuses on applying speech enhancement to observe the impact on real-world data. We also consider a baseline audio system to compare to our audio systems. Next, we discuss the details of the developed audio systems for 2019 NIST multimedia SRE.

\subsection{Baseline}

We use the standard Kaldi\footnotemark[2] system for having a baseline system for the audio track. The pre-trained x-vector model\footnotemark[3] is used to obtain a baseline system results. It is a system that uses wideband 16 kHz audio data for training the models and hence, we refer this system as x-vector$_{{\text {16k}}}$-I in this work.

\footnotetext[2]{http://kaldi-asr.org/}
\footnotetext[3]{http://kaldi-asr.org/models/m8}

\subsection{Mixed Bandwidth Systems}

Motivated by the gains achieved by mixed bandwidth systems in~\cite{NEC-TT2018CSL}, we are interested in studying the impact of narrowband and wideband systems to use their different aspects. This also leverage us to use larger amount of corpora for data augmentation during the training of x-vector extractor than that is used for the baseline.

We develop two systems with narrowband and wideband audio data, respectively. Both systems are composed by three modules, namely, feature extraction, x-vector extraction and linear discriminant analysis (LDA) with probabilistic LDA (PLDA). The difference of these two systems is on the sampling rate of speech data. The former considers 8 kHz audio data, whereas the latter considers 16 kHz audio data. Therefore, we refer these two systems as x-vector$_{\text {8k}}$ and x-vector$_{\text{16k}}$-II, respectively. The details of experimental setups for the two systems are mentioned in the following.

\subsubsection{Feature extraction}

23-dimensional mel frequency cepstral coefficient (MFCC) features are extracted from each utterance, which is followed by cepstral mean normalization with a window size of 3 seconds~\cite{Davis1980,Furui1981}. An energy based voice activity detection (VAD) method is then used to remove the silence frames for training and development test data. 

\subsubsection{x-vector extraction} 
\label{sec:vec_des}

The setting of x-vector network architecture (e.g., the number of hidden units per layer) and the training process follows that in~\cite{david_multiSpk}. The summary of corpora used for training, development and data augmentation can be viewed from Table~\ref{tab:data_part}. In addition, the specific details of the corpus used for training the extractor for narrowband and wideband systems are mentioned in Table~\ref{tab:tr_data}. We note that both narrowband and wideband systems follow the same network architecture to derive 512-dimensional x-vectors.

\begin{table}[t]
  \caption{Summary of speech corpora partitioned into training and development set.}
  \label{tab:data_part}
  \centering
              \begin{tabular}{l|l}
                \hline
                 \textbf{Corpus} & \textbf{Usage}  \\ \hline\hline
                 Previous NIST SREs & Training \\
                 Switchboard, VoxCeleb~\cite{Voxceleb,Voxceleb2} &  \\ \hline
                 MUSAN Music~\cite{MUSAN} \& Noise & Data Augmentation \\ \hline
                 RIRS Noise~\cite{RIRS} & Data Augmentation \\ \hline
                 2019 NIST SRE Dev & Development \\ \hline
              \end{tabular}
\end{table}

\begin{table}[t!]
  \caption{Details of training sets for x-vector extractor and LDA-PLDA model in narrowband and wideband systems.}
  \label{tab:tr_data}
  \centering
              \begin{tabular}{|c|c|}
                \hline
                 \textbf{System} & \textbf{Details}  \\ \hline\hline
                 \multicolumn{2}{|c|}{\bf x-vector extractor}\\ \hline
                 & 2,412,648 utterances from 11,577\\
                 Narrowband   & speakers in switchboard, previous SREs,\\
                 & VoxCeleb1 and VoxCeleb2 corpora. \\ \hline
                 Wideband   & 2,090,300 utterances from 7,185 speakers\\
                 &in VoxCeleb1 and VoxCeleb2 corpora. \\ \hline
                 \hline
                 \multicolumn{2}{|c|}{\bf LDA-PLDA}\\ \hline
                  Narrowband &176,081 utterances from 4,392 speakers\\
                  &  in previous SREs\\ \hline
                  Wideband   & 200,000 utterances from 6,345 speakers\\
                   &in VoxCeleb1 and VoxCeleb2 corpora\\ \hline
              \end{tabular}
  
\end{table}

\subsubsection{LDA-PLDA}

After pre-processing on the input vectors, LDA is first applied to minimize the channel and distance variation and reduce the dimension of x-vectors to 150. Then, PLDA is utilized as the back-end classifier. The data used for LDA-PLDA training for both narrowband and wideband systems can be viewed from Table~\ref{tab:tr_data}.

\subsection{WPE System}

The challenge data for 2019 NIST multimedia SRE are collected in uncontrolled environments and therefore, the audio segments are affected by background noise. Hence, we focused on speech enhancement aspect to improve the quality of input audio to the x-vector based system. The weighted prediction error (WPE) based dereverberation method is applied on the audio to perform speech enhancement~\cite{WPE_paper}. It is a statistical method that uses delayed linear prediction model to find out the prediction error, which is then optimized to obtain the clean speech~\cite{WPE_paper}. It is to be noted that we implement WPE only for the narrowband system for this study. The remaining system architecture follows the same as that of the x-vector$_{\text{8k}}$ system. We refer this WPE based system as x-vector$_{\text{8k}}$-WPE in this work. Next, we discuss the visual systems developed for 2019 NIST multimedia SRE.


\section{Visual Systems}
\label{seciii}

\begin{figure*}[t]
\begin{center}
    \includegraphics[width=1\linewidth]{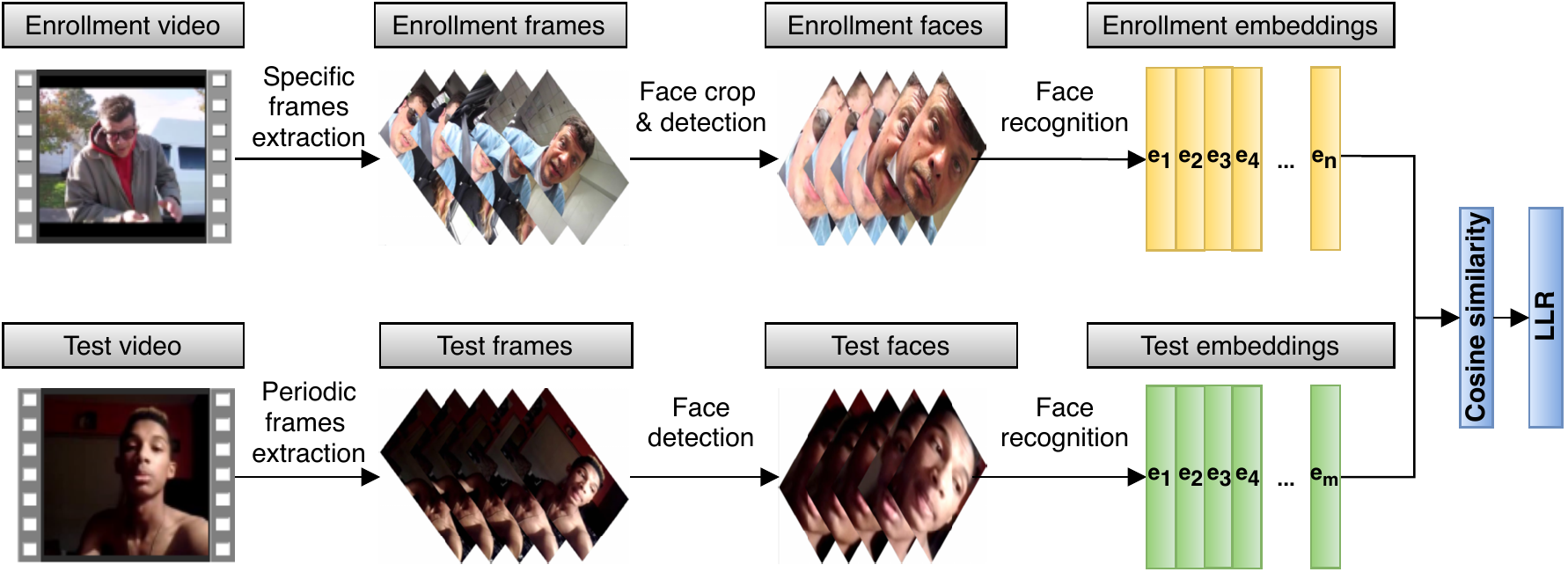}
	\end{center}
	\caption{{Block diagram of the visual system for 2019 NIST multimedia SRE. In the enrollment videos, the baseline method crops the target faces based on the given face bounding boxes, whereas the post evaluation method detects faces in the specific frames.}}
	\label{fig:VISUAL_BLOCK}
\end{figure*}

As discussed earlier, the 2019 NIST multimedia SRE involves both audio-visual cues for recognizing a person. Therefore, the enrollment video not only provides the target individual's voice segments, but also gives the target speaker's face information using the bounding boxes~\cite{sadjadiSRE2019}. The visual system is expected to automatically determine whether the target person is present in another test video. In this work, we develop two different systems for face recognition that considers visual information to recognize the speakers. The first one is based on ResNet-101 architecture and serves as a baseline system for visual modality. On the other hand, the second one uses the latest InsightFace based face embedding framework, which we develop during post evaluation. Table~\ref{tab:visual_data_part} summarizes various corpora that we used for developing the visual systems. We discuss the details of the baseline and the post evaluation visual systems in the following subsections.

\begin{table}[t]
  \caption{Summary of visual corpora partitioned into training and development set.}
  \label{tab:visual_data_part}
  \centering
  \begin{tabular}{l|l|l}
    \hline
     \textbf{Corpus} & \textbf{Usage} & \textbf{System} \\ \hline\hline
     MS-Celeb-1M \& & \multirow{2}{*}{Train Face Recognition} & \multirow{2}{*}{Baseline}\\
     Asian-Celebrity~\cite{megapixels} & & \\ \hline
     WIDER FACE~\cite{yang2016wider} & Train Face Detection & Post Evaluation \\ \hline
     MS1MV2~\cite{guo2016ms} & Train Face Recognition & Post Evaluation\\ \hline
     2019 NIST SRE Dev & Development & Both\\ \hline
  \end{tabular}
\end{table}

\subsection{Baseline: ResNet-101} 

In this subsection, we describe the baseline for the visual system that performs face recognition. It consists of a pipeline that receives raw images and produces the similarity scores between two segments. In the first stage, the specific faces in each enrollment video are cropped based on the given bounding boxes, then the frames in the whole test video at a rate of one frame per second are extracted. The face detection module takes the raw frames as input to detect the faces appeared. The detection contains a bounding box and five key points (eyes, nose and mouth), which is used to align the faces with similarity transformation. In the second stage, the aligned face images are converted to feature vectors with the recognition model. We use ResNet-101 trained on the cleaned version of MS-Celeb-1M database and Asian-Celebrity database as the recognition model~\cite{megapixels}. In the third stage, we perform template matching operation to match faces between two segments. The pair-wise similarity scores are computed based on the extracted face embeddings, and the average of top-10 pairs is regarded as the final score. Fig.~\ref{fig:VISUAL_BLOCK} shows the block diagram of the visual system with different stages involved to perform face recognition.

\subsection{Post Evaluation: InsightFace}

During post evaluation, we primarily focused on improving our visual system. The general pipeline of the system follows that of the baseline, but we used the latest methods at various stages for face recognition~\cite{cross_modal_is2020}. We highlight the key differences of the post evaluation visual systems in the following. 

\subsubsection{Face Detection} 

Face detection is performed with RetinaFace~\cite{deng2019retinaface}, which is the state-of-the-art framework in WIDER FACE hard test set~\cite{yang2016wider}. The RetinaFace is a robust single-stage face detector, which added five facial landmarks as the extra supervision signal to improve the performance. We selected ResNet-50 model, which is trained on the WIDER FACE database for face detection. Followed by this, multi-task cascaded convolutional networks (MTCNN)~\cite{zhang2016joint} is applied to perform face alignment. In this study, for enrollment videos, we found the quality of some given bounding boxes are not so high due to the poses, lighting conditions and crop quality, hence, instead of using the given frames, 5 frames (given frames plus the $ \pm $ 2 frames for each one) are selected as the specific frames to improve the performance. For these frames, we extracted and detected the faces, whose overlap rate with the given boxes is higher than the threshold. 

\subsubsection{Face Embedding}

The face embeddings are extracted using InsightFace~\cite{deng2019arcface}, which considers additive angular margin loss to obtain highly discriminative features for face recognition. This framework can find the maximum classification boundary in the angular space. We considered ResNet-100 model trained on the cleaned MS-Celeb-1M database (MS1MV2)~\cite{guo2016ms} to obtain 512-dimensional face embeddings. 

\subsubsection{Backends}

For each trial, the cosine similarity score between the extracted face embeddings in the enrollment video and the test video is computed. Instead of selecting the top-10 scores as that in the baseline method, the average of top 20\% scores is taken here to get the final log-likelihood ratio (LLR) result using logistic regression.


\section{Results and Analysis}
\label{seciv}

In this section, we report the results and analysis of different systems developed for 2019 NIST multimedia SRE. We used the scoring tool package given by the organizers to report the results in terms of equal error rate (EER), minimum detection cost function (minDCF) and actual detection cost function (actDCF) as per the evaluation plan~\cite{sre19_plan}. It is noted that actDCF serves as the primary metric for the challenge. The score level fusion and calibration of various systems are performed using the Bosaris\footnotemark[4] toolkit~\cite{bosaris} in this study.

\begin{table} [t!]
\caption{\label{table_results} {Performance of audio, visual and audio-visual (AV) systems on the development set of 2019 NIST multimedia SRE. }}
\centerline{
\begin{tabular}{|c|c||c|c|c|}
\hline
{\bf Modality} & {\bf System} &  {\bf EER (\%)} & {\bf minDCF}& {\bf actDCF}\\
\hline
\hline
 & x-vector$_{{\text {16k}}}$-I & 12.96 & 0.525 & 0.548\\
 & x-vector$_{\text {16k}}$-II &  12.38 & 0.463 & 0.838 \\
Audio &x-vector$_{\text{8k}}$ &  12.04 & 0.467 & 0.632 \\
 & x-vector$_{\text{8k}}$-WPE & 11.42 & 0.462 & 0.638 \\
 \cline{2-5}
 & Fusion: Audio & 12.04 & 0.459 &{0.466}\\ \hline \hline
 Visual &ResNet-101 & {11.11}&{ 0.423} &{ 0.446}\\
 \hline
\hline
\multicolumn{5}{|c|}{\bf Challenge Submission}\\
\hline
Audio-visual & Fusion: AV &{\bf 6.48} &{\bf 0.208} &{\bf 0.216}\\ \hline\hline
\multicolumn{5}{|c|}{\bf Post Evaluation}\\
\hline
Visual & InsightFace & 5.65 & {0.318} & {0.342}\\\hline\hline
Audio-visual & Fusion: AV& {\bf 4.63} &{\bf 0.130} &{\bf 0.157}\\\hline
\end{tabular}}
\vspace{-4mm}
\end{table}

\footnotetext[4]{https://sites.google.com/site/bosaristoolkit/}

Table~\ref{table_results} shows the results for various audio and visual systems on the development set of 2019 NIST multimedia SRE. We note that scores belonging to the individual audio or visual systems are not calibrated. The scores are only calibrated for the fusion systems. It is observed that our mixed bandwidth and WPE based systems perform better than the baseline audio system. Further, the score level fusion of the audio systems improves the actDCF as well as minDCF compared to all the individual audio systems. This shows the gain achieved by complementary nature of information by these developed audio systems.  

We then focus on the results of the visual systems. It is observed from Table~\ref{table_results} that the visual system performs better than all the audio systems as well as fusion of all the audio systems, which shows its effectiveness for recognizing speakers. Further, we perform the fusion of all the four audio systems and the ResNet-101 based visual system, which was our primary submission to the audio-visual track of 2019 NIST multimedia SRE. We find the fused audio-visual system enhances the performance by a larger margin due to the complementary information for capturing speaker identity.

Post evaluation, we developed another visual system based on InsightFace framework, whose results are also reported in Table~\ref{table_results}. This system performs even better than our baseline ResNet-101 based visual system that was used for audio-visual fusion during challenge submission. The gains achieved with this post evaluation system is due to the use of the state-of-the-art face detection as well as face recognition models. Further, the scoring strategy described in the previous section also helped to obtain an improved result. We then perform a fusion of this system with the four audio and ResNet-101 based visual systems. The resultant fused system obtained post evaluation significantly improves the performance of audio-visual speaker recognition.

Table~\ref{table_results_eval} shows the results of various systems and their fusion on the evaluation set of 2019 NIST multimedia SRE. We observe that the performance of all the audio and visual systems are comparatively better than that on the development set. This points towards the fact that the development set is more challenging in nature. Further, the WPE based audio system emerges as the best system among the all audio system developed on the evaluation set. On the other hand, the InsightFace system outperforms the baseline visual system by a large margin. This also results in a significant gain in the audio-visual fusion system obtained with post evaluation study. We obtain an actDCF of 0.026 with the final audio-visual system in comparison to actDF of 0.136 audio-visual system submitted to the challenge. It is also noted that the final audio-visual system thus obtained performs comparable to the top two systems of 2019 NIST multimedia SRE.

\begin{table} [t!]
\caption{\label{table_results_eval} {Performance of audio, visual and audio-visual (AV) systems on the evaluation set of 2019 NIST multimedia SRE. }}
\centerline{
\begin{tabular}{|c|c||c|c|c|}
\hline
{\bf Modality} & {\bf System} &  {\bf EER (\%)} & {\bf minDCF}& {\bf actDCF}\\
\hline
\hline
 & x-vector$_{{\text {16k}}}$-I & 7.79 & 0.407 & 0.610\\
 & x-vector$_{\text {16k}}$-II & 7.30 & 0.317 & 0.558 \\
Audio & x-vector$_{\text{8k}}$ & 8.14 & 0.414 & 0.548 \\
 & x-vector$_{\text{8k}}$-WPE & 7.02 & 0.306 & 0.529 \\
 \cline{2-5}
 & Fusion: Audio & 6.46 &0.314 & 0.321\\ \hline \hline
 Visual & ResNet-101 & 6.16 & 0.330 &	0.343\\
 \hline
\hline
\multicolumn{5}{|c|}{\bf Challenge Submission}\\
\hline
Audio-visual & Fusion: AV &{\bf 2.32} &	{\bf 0.131} &	{\bf 0.136}\\ \hline\hline
\multicolumn{5}{|c|}{\bf Post Evaluation}\\
\hline
 Visual & InsightFace & 1.55 & 0.049 & 0.085\\\hline\hline
Audio-visual & Fusion: AV& {\bf 0.88} &	{\bf 0.024} &	{\bf 0.026}\\\hline
\end{tabular}}
\vspace{-4mm}
\end{table}

%
%

\section{Conclusion}
\label{conc}

This work reports the details of audio-visual systems developed by HLT-NUS for 2019 NIST multimedia SRE. The audio systems follow x-vector based framework with the focus on mixed bandwidth and speech enhancement. On the contrary, the visual systems are developed using ResNet-101 and InsightFace frameworks. A score level fusion of the developed audio and visual systems significantly improves the speaker recognition performance in comparison to systems based on individual modality. The future work will focus on associating temporal relation of audio-visual cues for performing multimedia speaker recognition.

\section{Acknowledgement}

This research work is supported by Programmatic Grant No. A1687b0033 from the Singapore Government's Research, Innovation and Enterprise 2020 plan (Advanced Manufacturing and Engineering domain), and Human-Robot Interaction Phase 1 (Grant No. 192 25 00054) from the National Research Foundation, Prime Minister's Office, Singapore under the National Robotics Programme.

\balance
\bibliographystyle{IEEEtran}

\bibliography{MyReferences_new}

\end{document}